\documentclass[twocolumn,prb,aps,showpacs,superscriptaddress]{revtex4-1}

\usepackage{graphicx}
\usepackage{amssymb}
\usepackage{amsfonts}
\usepackage[T1]{fontenc}
\usepackage[latin1]{inputenc}

\usepackage{graphicx}
\usepackage{psfrag}
\usepackage{color}
\usepackage{amsfonts}
\usepackage{amsmath}
\usepackage{amssymb}
\usepackage{ifthen}
\usepackage[latin1]{inputenc}
\usepackage{bm}

\definecolor{hellgelb}{cmyk}{0, 0.1, 0.7, 0}
\definecolor{OliveGreen}{cmyk}{0.40,0,0.95,0.25}
\definecolor{DarkGreen}{cmyk}{0.70,0,1,0.45}
\definecolor{hellrot}{cmyk}{0,0.3,0.4,0}
\definecolor{hellgruen}{cmyk}{0.25,0,0.45,0}

\begin{document}

\title{Toolbox of resonant quantum gates in circuit quantum electrodynamics}


\author{G. Haack}
\affiliation{D\'epartement de Physique Th\'eorique, Universit\'e de Gen\`eve, CH-1211 Gen\`eve 4, Switzerland}

\author{F. Helmer}
\affiliation{Department of Physics, ASC, and CeNS, Ludwig-Maximilians-Universit\"{a}t, Theresienstrasse 37, 80333
M\"{u}nchen, Germany}

\author{M. Mariantoni}
\affiliation{Department of Physics, University of California, Santa Barbara, CA 93106, USA}

\author{F. Marquardt}
\affiliation{Institut f\"ur Theoretische Physik, Universit\"at Erlangen-N\"urnberg, Staudtstr. 7, 91058 Erlangen, Germany}

\author{E.~Solano}
\affiliation{Departamento de Qu\'{\i}mica F\'{\i}sica, Universidad del Pa\'{\i}s Vasco -
Euskal Herriko Unibertsitatea, Apdo. 644, 48080 Bilbao, Spain}
\affiliation{IKERBASQUE, Basque Foundation for Science, Alameda Urquijo 36, 48011 Bilbao,
Spain}
\date{\today}


\begin{abstract}
We propose the implementation of fast resonant gates in circuit quantum electrodynamics for
quantum information processing. We show how a suitable utilization of three-level
superconducting qubits inside a resonator constitutes a key tool to perform diverse
two-qubit resonant gates, improving the operation speed when compared to slower dispersive
techniques. To illustrate the benefit of resonant two-qubit gates in circuit QED, we consider the implementation of a two-dimensional cluster state in an array of $N \times N$ superconducting qubits by using resonant controlled-phase (CPHASE) and one-qubit gates, where the generation time grows linearly with $N$. For $N=3$, and taking into account decoherence mechanisms, a fidelity over $60\%$ for the generation of this cluster state is obtained.
\end{abstract}

\maketitle

\section{Introduction}

Circuit quantum electrodynamics (QED) is a novel field combining atomic physics and
quantum optical cavity QED concepts with superconducting circuits.\cite{EarlyCircuitQED,Blais04,ChiorescuWallraff} Its fundamental dynamics is
understood within the Jaynes-Cummings model, describing the interaction between a
two-level system and a single field mode.\cite{Raimond01} In circuit QED,
superconducting qubits are considered as artificial atoms \cite{Makhlin01,Qubitpapers}
interacting with on-chip one-dimensional resonators playing the role of cavities.\cite{SchoelkopfGirvin} These mesoscopic devices are candidates for implementations in quantum information processing~\cite{Nielsen04} due to their inherent tunability and
scalability properties, at least as good as trapped ions~\cite{Leibfried03} or quantum dots.\cite{Hanson07}

To implement quantum algorithms in circuit QED\cite{DiCarlo09} within the standard
quantum computing approach, the sequential realization of fast high-fidelity quantum
gates is required.\cite{Blais07} Several implementations of quantum operations in
coupled superconducting qubits have been proposed\cite{Schuch03,Galiautdinov09} and the
implementation of a controlled-NOT gate has been achieved.\cite{Plantenberg07}
Furthermore, coherent coupling\cite{Majer07} and quantum information
exchange\cite{Sillanpaa07} of two superconducting qubits, through a cavity bus and cavity
state synthesis\cite{Hofheinz09}, have been realized. Scalable architectures in circuit
QED may require efficient designs of two-dimensional cavity arrays.\cite{Helmer09} These allow scalable standard quantum
computation, and also open the possible realization of two-dimensional cluster states
for one-way quantum computing.\cite{Raussendorf01,Tanamoto09} Another crucial challenge in quantum information and also in circuit QED is to speed up operations and, thus, beat decoherence. For this purpose it would be desirable to employ resonant gates. Indeed, based on first-order couplings between qubits and cavities, resonant gates are much faster than the commonly used dispersive gates, based on second-order couplings. In cavity QED, a controlled-phase (CPHASE) gate has been already implemented in the resonant regime~\cite{Rauschenbeutel99}, but these ideas are difficult to scale up and build, for example, cluster states in multiqubit systems. In this work, we focus on the field of circuit QED and propose the implementation of a toolbox of resonant quantum gates
improving on speed and fidelity when compared to slower dispersive gates in circuit QED. To this end, a key concept from quantum optics  will be the use of an auxiliary excited state for the qubits and the use of the cavity as a resonant mediator of the qubit interactions.

The paper is organized as follows: in Section II, we will explain how to realize, resonantly and efficiently, the paradigmatic two-qubit CPHASE gate. The theoretical protocol is inspired by cavity QED experimental works\cite{Nogues99}. To prove its feasibility in circuit QED, we test our results with full numerical simulations involving decoherence mechanisms, reaching a fidelity of $98.5\%$. Then, in Section III, we extend the ideas of previous section and propose the implementation of other two-qubit gates: the iSWAP gate and the Bogoliubov gate.~\cite{Verstraete09} Some of these two-qubit gates, together with suitable
one-qubit gates, can form universal sets for quantum computation. In Section IV, as an example of the resonant CPHASE gate presented before, we propose the realization in circuit QED of a cluster state for one-way quantum
computing\cite{Raussendorf01} in a two-dimensional array of cavities.~\cite{Helmer09}

\section{Model for a resonant CPHASE gate}

The CPHASE gate is one of the paradigmatic two-qubit gates for quantum information. Indeed, by combining it with one-qubit gates, it forms a set of universal gates. This CPHASE gate produces a phase-shift $\pi$ only when both qubits are excited. In the computational basis of two qubits, $\{ \vert \rm{g}_1 \rm{g}_2\rangle, \vert \rm{g}_1 e_2\rangle, \vert e_1 \rm{g}_2\rangle, \vert e_1 e_2\rangle \}$, the CPHASE matrix reads
\begin{displaymath}
\text{CPHASE} = \left( \begin{array}{cccc}
1 & 0 & 0 &0 \\
0 & 1 & 0 &0 \\
0 & 0 & 1 &0 \\
0 & 0 & 0 &-1
\end{array} \right),
\end{displaymath}
where $| \rm{g} \rangle$ and $| \rm{e} \rangle$ represent the ground and excited states of the qubits, labeled with subindices $1$ and $2$. In a recent experiment~\cite{DiCarlo09}, where a quantum algorithm was implemented in circuit QED, a dispersive interaction using the third level of a superconducting qubit was used. To implement a CPHASE gate in the resonant regime, in which the detuning is zero, we borrow from quantum optics the use of the cavity as a mediator for the operation and the use of auxiliary qubit levels.\cite{Nogues99} We develop below a specific modelisation in the context of circuit QED and present numerical simulations, including realistic losses, of the involved superconducting qubits and microwave resonators.

We consider two three-level superconducting qubits, 1 and 2, with $| {\rm g}_{1,2}
\rangle$, $| {\rm e}_{1,2} \rangle$, and $| {\rm a}_{1,2} \rangle$ being their first
three lower energy levels, respectively. The usefulness of auxiliary qubit levels in the
superconducting qubit context has been recognized only recently, for phase
qubit operations,\cite{Strauch03, Sillanpaa09} for circuit QED quantum optics
applications,\cite{Marquardt07} as well as for a recent demonstration of gates in circuit
QED.\cite{DiCarlo09} The qubits are coupled to a coplanar waveguide cavity and their dynamics is described,
after a rotating-wave approximation, by the Hamiltonian
\begin{eqnarray}
H & = & \sum_{l=g,e,a;q=1,2} E_{l_q} \left|l_q\right>\left<l_q\right|  + \hbar\omega_{\rm
r} a^{\dagger} a
  \nonumber\\
 &&
   + \hbar g_{\rm g_1e_1} \left( \sigma_{\rm g_1e_1}^+ a
     + \sigma_{\rm g_1e_1}^- a^{\dagger}\right)      \nonumber\\
 &&  + \hbar g_{\rm e_1 a_1} \left( \sigma_{\rm e_1 a_1}^+ a
     + \sigma_{\rm e_1 a_1}^- a^{\dagger}\right)     \nonumber\\ &&
+ \hbar g_{\rm g_2 e_2} \left( \sigma_{\rm g_2 e_2}^+ a
     + \sigma_{\rm g_2 e_2}^- a^{\dagger}\right) \nonumber\\
 &&
   + \hbar g_{\rm e_2 a_2} \left( \sigma_{\rm e_2 a_2}^+ a
     + \sigma_{\rm e_2 a_2}^- a^{\dagger}\right)     .
\label{hamsimple}
\end{eqnarray}
Here, $\sigma^+_{k, l} \equiv |
{l} \rangle \langle {k} |$, $\sigma^-_{k l} \equiv |
{k} \rangle \langle  {l} |$, while $a\:(a^{\dagger})$ are the bosonic
annihilation (creation) operators of the resonator field mode, $E_{l_q}$ is the energy
for level $l$ of qubit $q$, $\omega_r$ is the cavity mode frequency, and the $g$'s denote
the vacuum Rabi coupling strengths. We assume that the qubit levels are anharmonic and
that each transition can be tuned to match the cavity frequency, e.g. by using
AC-Stark-shift fields,\cite{Schuster05} thereby effectively switching on the
qubit-cavity couplings for the respective transition. Conversely, we could make use of a tunable cavity.\cite{Sandberg08}

We assume that at a certain point of a quantum computation the tripartite system (qubits+cavity) is in the following state
\begin{eqnarray}
\vert \Psi_i \rangle \! = \!\! \Big( \alpha \, \vert \rm{g}_1 \rm{g}_2\rangle + \beta \,
\vert \rm{g}_1 e_2\rangle + \gamma \, \vert e_1 \rm{g}_2\rangle + \delta \, \vert e_1
e_2\rangle \Big) \! \otimes \! | 0 \rangle ,
\end{eqnarray}
where $\alpha, \beta, \gamma$ and $\delta$ are arbitrary complex amplitudes and $| 0 \rangle$
is the cavity vacuum. A CPHASE gate is implemented on qubits 1 and 2 after the protocol
displayed in Fig.~1 is applied. In step (i), the state of qubit 2, encoded in its two lowest
energy levels, $| {\rm g}_2 \rangle$ and $| {\rm e}_2 \rangle$, is mapped onto a photonic
cavity qubit through a resonant coupling between qubit 2
and the cavity mode, while the first qubit remains off-resonant.
It is possible to achieve this mapping without introducing extra phase factors, such that the state at that stage is $\vert \Psi \rangle \! = \!\! ( \alpha \, \vert
\rm{g}_1 0 \rangle + \beta \, \vert \rm{g}_1 1 \rangle + \gamma \, \vert e_1 0 \rangle +
\delta \, \vert e_1 1 \rangle ) \otimes | \rm{g}_2 \rangle$. In step
(ii)~\cite{Nogues99}, a 2$\pi$ resonant pulse between the two upper energy levels of
qubit 1, $| {\rm e}_1 \rangle$ and $| {\rm a}_1 \rangle$, and the cavity mode realizes a
CPHASE gate among them, yielding $\vert \Psi \rangle \! = \!\! ( \alpha \, \vert \rm{g}_1
0 \rangle + \beta \, \vert \rm{g}_1 1 \rangle + \gamma \, \vert e_1 0 \rangle - \delta \,
\vert e_1 1 \rangle ) \otimes | \rm{g}_2 \rangle$. In the last step (iii), the cavity
qubit is mapped back to qubit 2, thus implementing an overall CPHASE gate between qubits
1 and 2, leaving the cavity qubit decoupled in the vacuum state
\begin{eqnarray}
\vert \Psi_f \rangle \! = \!\! \Big( \alpha \, \vert \rm{g}_1 \rm{g}_2\rangle + \beta \,
\vert \rm{g}_1 e_2\rangle + \gamma \, \vert e_1 \rm{g}_2\rangle - \delta \, \vert e_1
e_2\rangle \Big) \! \otimes \! | 0 \rangle .
\end{eqnarray}

\begin{figure}[t]
\includegraphics[width=7cm]{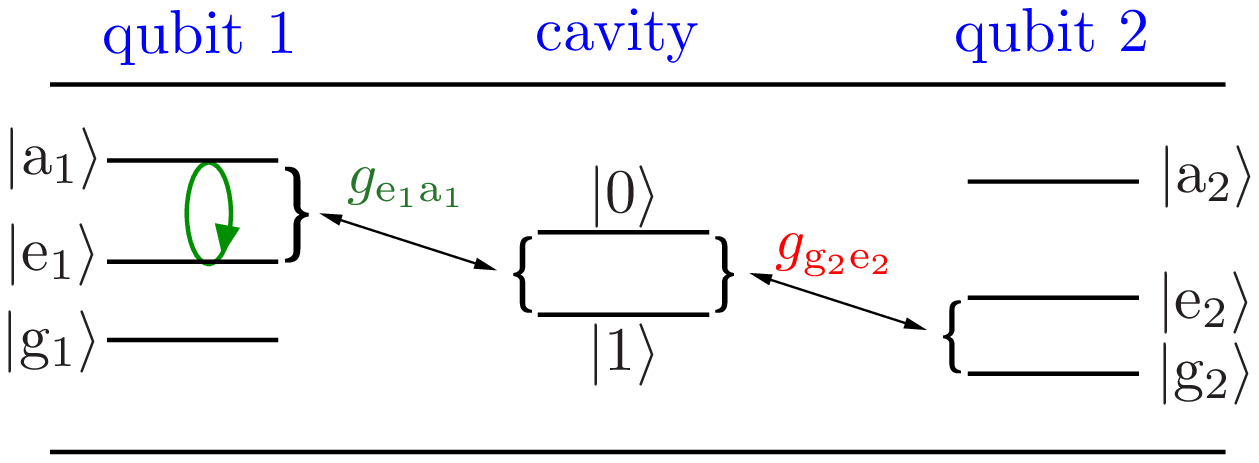}
\begin{center}

\begin{tabular}{|l|c|c|c|}
\hline
\text{Step} & \text{Transition} & \text{Coupling} & \text{Pulse}\\ \hline
\text{i) Map}& $\vert \text{g}_2 \rangle, \vert \text{e}_2\rangle \rightarrow \vert 0
\rangle, \vert1 \rangle$&$g_{\rm g_2 e_2} \color{black}(2\pi \cdot 100
\text{MHz})$  & $\pi$ \\ \hline
\text{ii) CPHASE}&  $\vert \text{e}_1 \rangle, \vert \text{a}_1\rangle \leftrightarrow
\vert 0 \rangle, \vert1 \rangle$& $g_{\rm e1 a1} \color{black}(2\pi
\cdot 38.8 \text{MHz})$ & $2\pi$ \\ \hline
\text{iii) Map back}& $\vert \text{g}_2 \rangle, \vert \text{e}_2\rangle \leftarrow \vert
0 \rangle, \vert1 \rangle$ & $g_{\rm g_2 e_2} \color{black}(2\pi \cdot 100
\text{MHz})$ & $\pi$\\ \hline
\end{tabular}
\end{center}
\caption{Sketch and protocol for the resonant implementation of a CPHASE gate between
qubits 1 and 2. The coupling values given in parentheses are the ones used for the
simulation presented in Fig.2.}
\end{figure}

To prove in circuit QED the feasibility of the proposed resonant protocol for the implementation of a
CPHASE gate, we employ a time-dependent Lindblad master equation simulation,
taking into account cavity losses and qubit relaxation and dephasing. We consider
a charge qubit where the ratio between the charge and the Josephson energies, $E_J /
E_C$, can be tuned from the charge to the transmon regime.\cite{Houck08} Even if the
transmon regime enjoys the best decoherence parameters, it is not obvious that this limit
will be optimal for having the required level anharmonicity when tuning different qubit
transitions to the cavity mode, or vice versa. Note that, in contrast to the case of
dispersive gates, where only virtual photons are involved, in step (ii) we populate
resonantly the cavity with a real photon during a finite operation time. The simulations
indicate that the anharmonicity condition for the qubit levels and the decay of cavity
photons do not prevent the generation of a fast high-fidelity CPHASE gate between the
qubits. As an initial state, we have chosen
\begin{eqnarray}
\vert \bar{\Psi}_i \rangle \! = \!\! \frac{1}{\sqrt{2}} \Big( \vert \rm{g}_1 g_2\rangle +
\vert e_1 \rm{e}_2\rangle \Big) \otimes | 0 \rangle ,
\end{eqnarray}
where the maximally entangled qubits should be a sensitive probe of decoherence
processes. Figure~2 shows the reduced density operator of the qubits at the initial and
final state, after the three steps to achieve the CPHASE gate. The values we
considered in the simulation are
$\kappa_{\text{cav}} = 10 \text{ kHz}$ for the cavity decay rate (with a cavity frequency
of $\omega_{\rm cav}=2\pi\cdot5.5 {\rm GHz}$), and $T_1 = 7.3 \mu s$
and $T_2 = 0.5 \mu s$ for the time-scales of qubit relaxation and decoherence,
respectively.
The coupling strengths have been obtained from diagonalizing the charge qubit
Hamiltonian for $E_J/E_C=2.53$ and are indicated in the table in Fig.\ 1. In the
simulation, level positions have been shifted during the time-evolution, assuming an
appropriate AC Stark shift, to realize the required qubit-resonator resonances. The density
operators are expressed in the computational basis $\{ \vert \rm{g}_1 \rm{g}_2\rangle,
\vert \rm{g}_1 e_2\rangle, \vert e_1 \rm{g}_2\rangle, \vert e_1 e_2\rangle \}$.
We find a final state fidelity of 98.5 \% in our simulation, where the fidelity is defined as $\text{F} = {\rm Tr} \big( \vert \sqrt{\rho_{\text{lossy}}}
\rho_{\text{ideal}} \sqrt{\rho_{\text{lossy}}} \vert \big)$, with $\rho_{\text{ideal}}$
and $\rho_{\text{lossy}}$ being the final state density operators for the ideal operation
and for the real evolution, respectively.

We want to emphasize here that this simulation for the charge qubit regime
only serves as an illustration for the large range of possible experimental
setups in circuit QED that may be employed to implement the resonant gate protocol suggested in this paper.
Optimization of parameters to overcome practical challenges (like fluctuating stray charges) is
of course mandatory but it is best left to any individual experimental design separately.

\begin{figure}[h!]
\centering
\includegraphics*[width=0.45\columnwidth]{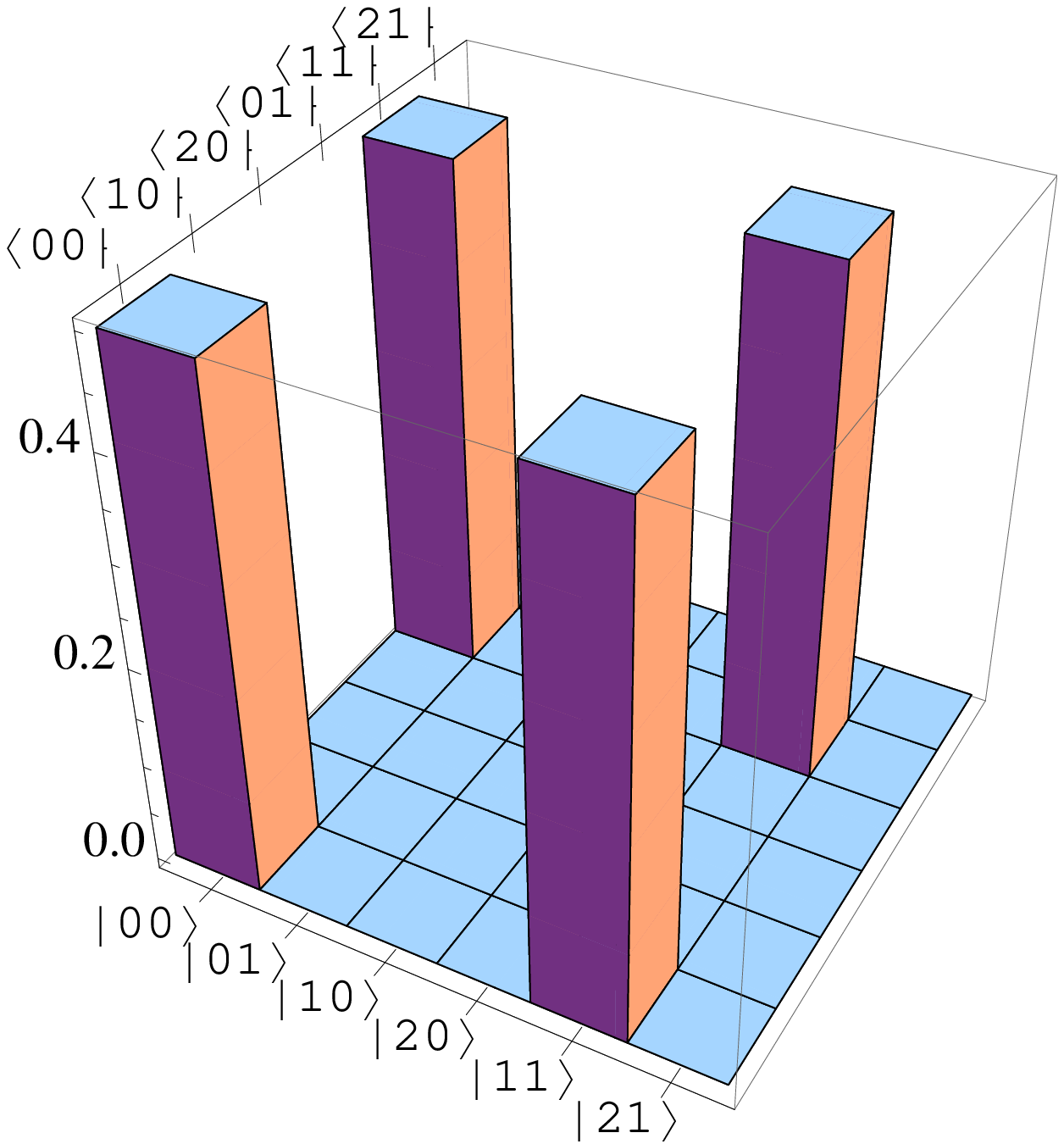}
\includegraphics*[width=0.45\columnwidth]{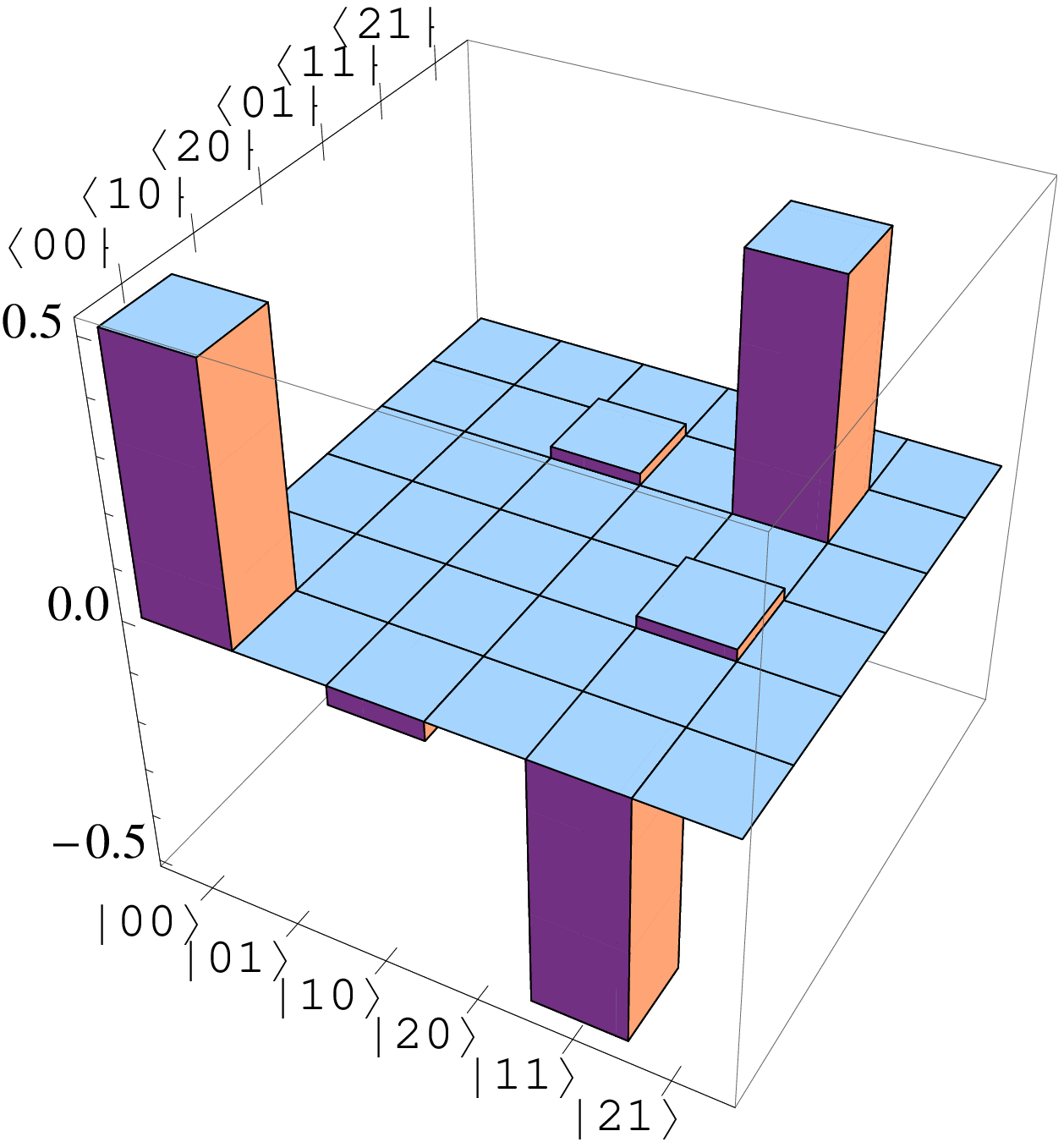}
\caption{Density operator of the initial state $\Psi_i$ (with $\alpha = \delta = 1, \beta
= \gamma =0$) and of the final state after the implementation of the CPHASE gate, in a
simulation involving qubit relaxation and dephasing, and cavity decay; expressed in the
computational basis of the two qubits.}
\label{simu_ini}
\end{figure}

In this section we have shown that this protocol using three-level physics and the cavity-resonator as mediator allows the implementation of a CPHASE gate between two superconducting qubits in circuit QED with high fidelity. According to this promising result, we will give in the next section the protocols for another important two-qubit gate, the iSWAP gate, and a more unusual two-qubit gate, the Bogoliubov one. Whereas the implementation of the CPHASE gate was inspired from cavity QED, the protocols for these two gates are new.

\section{Toolbox of resonant gates}

Most two-qubit gates form a universal set if properly accompanied by the suitable
one-qubit gate. In this sense, it should be enough to have a two-qubit gate, say the
CPHASE, that can be done fast and efficiently. However, depending on the quantum
computation one wants to implement, a more diverse gate toolbox leads to a more efficient
protocol. We show now that the proposed resonant tools are not only valid for generating
resonantly and efficiently the CPHASE gate, but can also produce some other classical and
exotic two-qubit gates. We exemplify this with the iSWAP and the Bogoliubov gate, see Fig.~3 and Fig.~4. The latter appears for instance
in the context of simulations of the dynamical evolution of spin models with quantum gates.\cite{Verstraete09}

The two-qubit iSWAP gate, a paradigmatic gate appearing in circuit QED, rotates with a given phase the second qubit if the first qubit is in its excited state. In the computational two-qubit basis its matrix reads
\begin{displaymath}
\text{iSWAP} = \left( \begin{array}{cccc}
1 & 0 & 0 &0 \\
0 & 0 & i &0 \\
0 & i & 0 &0 \\
0 & 0 & 0 &1
\end{array} \right).
\end{displaymath}
The standard approach to generate the iSWAP gate is based on a weak and slow second-order coupling, according to the flip-flop interaction Hamiltonian for which this two-qubit gate occurs naturally.~\cite{Schuch03, Blais07} The iSWAP is considered as a paradigmatic gate in the context of circuit QED, but it is known to be a slow dispersive two-qubit gate. The protocol in Fig.~3 presents the different steps to implement this gate in the resonant regime. Compared to the CPHASE gate, one more step is required for rotating the ground state of the first qubit according to the definition of the iSWAP gate. For this, one has to shelve the population of its excited state, step (ii) in Fig.~3.

\begin{figure}[t]  \label{fig3}
\includegraphics[width=7cm]{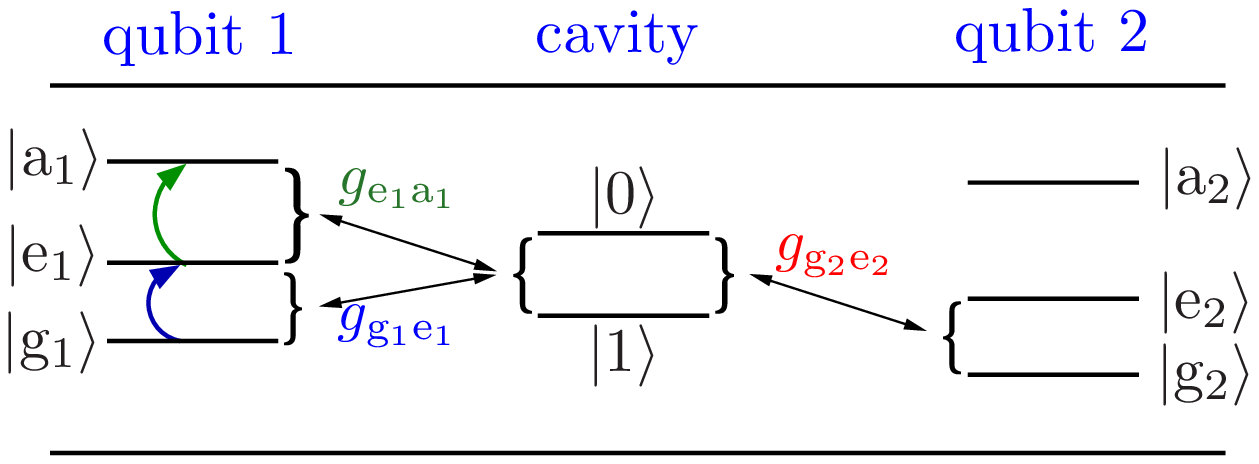}
\begin{center}

\begin{tabular}{|l|c|c|c|}
\hline
\text{Step} & \text{Transition} & \text{Coupling} & \text{Pulse}\\ \hline
\text{i) Mapping}& $\vert \text{g}_2 \rangle, \vert \text{e}_2\rangle \rightarrow \vert 0
\rangle, \vert1 \rangle$&$g_{\text{g}_2\text{e}_2}$ & $\pi$ \\ \hline
\text{ii) Shelving}&  $\vert \text{e}_1 \rangle, \vert \text{a}_1\rangle \leftrightarrow
\vert 0 \rangle, \vert1 \rangle$&$g_{\text{e}_1\text{a}_1}$ & $\pi$ \\
\hline
\text{iii) Rotate qubit 1}& $\vert \text{g}_1 \rangle, \vert \text{e}_1\rangle
\leftrightarrow \vert 0 \rangle, \vert1 \rangle$&$g_{\text{g}_1\text{e}_1}$ &
$\pi$ \\ \hline
\text{iv) Back shelving}&  $\vert \text{e}_1 \rangle, \vert \text{a}_1\rangle
\leftrightarrow \vert 0 \rangle, \vert1 \rangle$&
$g_{\text{e}_1\text{a}_1}$ & $-\pi$ \\ \hline
\text{v) Back mapping}& $\vert \text{g}_2 \rangle, \vert \text{e}_2\rangle \leftarrow
\vert 0 \rangle, \vert1 \rangle$ &$g_{\text{g}_2\text{e}_2}$ & $\pi$ \\ \hline
\end{tabular}
\caption{Sketch and protocol for the resonant implementation of an iSWAP gate between
qubits 1 and 2.}
\end{center}
\end{figure}

As last example for the use of our method, we consider the Bogoliubov gate. This two-qubit gate is not common and plays a role for instance in the simulation of the Ising model in an arbitrary transverse field.\cite{Verstraete09} In the computational basis of two qubits, its matrix form reads
\begin{displaymath}
\text{B} = \left( \begin{array}{cccc}
\cos \theta & 0 & 0 &i \sin \theta \\
0 & 1 & 0 &0 \\
0 & 0 & 1 &0 \\
i \sin \theta & 0 & 0 & \cos \theta
\end{array} \right),
\end{displaymath}
where the angle $\theta$ depends on the external transverse magnetic field $\lambda$, $\theta = \arctan (\lambda - \sqrt{1+\lambda^2})$. The protocol is shown in Fig.\ 4. More steps are required to implement this gate, but all of them are in the resonant regime, showing that our tools are universal and could be applied to different kinds of two-qubit unitary transformations.

It is also desirable that the one-qubit gates needed to form a universal set of gates can be implemented with fast resonant pulses. As it is well known, rotations around the x and y axis can always be
implemented by sending an external driving field resonant to the transition frequency of
the corresponding qubit. Rotations around the z axis correspond to a detuning, which for
charge qubits can be implemented via changing the Josephson energy $E_J$ or by AC Stark
shifts. Employing AC Stark shifts has the advantage of potentially being able to get rid of fast control pulses on a separate control line, which would be needed for tuning $E_J$. An AC Stark shift is implemented by irradiating the qubit via the cavity. The microwave drive, being detuned from the qubit, then generates a shift of the qubit transition frequency, leading to an additional phase shift between the two qubit levels. In particular, this allows to implement the Hadamard gate, either by a 180
degrees rotation around $x+z$ or by the sequence $H = e^{i\pi/2}R_x(\pi)R_y(\pi/2)$,
where $R_i(\theta)$ is a rotation by $\theta$ around axis $i$. Together with CPHASE (or
CNOT) and a $\pi/8$ gate, this forms a universal set of gates for quantum computing.

\begin{figure}
\begin{tabular}{|l|c|c|c|}
\hline
\text{Step} & \text{Transition} & \text{Coupling} & \text{Pulse}\\ \hline
\text{i) Mapping}& $\vert \text{g}_2 \rangle, \vert \text{e}_2\rangle \rightarrow \vert 0
\rangle, \vert1 \rangle$&$g_{\text{g}_2\text{e}_2}$ & $\pi$ \\ \hline
\text{ii) Rotate qubit 1}& $\vert \text{g}_1 \rangle \leftrightarrow \vert
\text{e}_1\rangle $&$\Omega_R$ & $\pi$ \\ \hline
\text{iii) Shelving}&  $\vert \text{e}_1 \rangle, \vert \text{a}_1\rangle \leftrightarrow
\vert 0 \rangle, \vert1 \rangle$&$g_{\text{e}_1\text{a}_1}$ & $\pi$ \\
\hline
\text{iv) Entanglement}& $\vert \text{g}_1 \rangle, \vert \text{e}_1\rangle
\leftrightarrow \vert 0 \rangle, \vert1 \rangle$&$g_{\text{g}_1\text{e}_1}$ &
$2\theta$ \\ \hline
\text{v) Back shelving}&  $\vert \text{e}_1 \rangle, \vert \text{a}_1\rangle
\leftrightarrow \vert 0 \rangle, \vert1 \rangle$&
\color{DarkGreen}$g_{\text{e}_1\text{a}_1}$ & $-\pi$ \\ \hline
\text{vi) Back rotation}& $\vert \text{g}_1 \rangle \leftrightarrow \vert
\text{e}_1\rangle $&$\Omega_R$ & $\pi$ \\ \hline
\text{vii) Back mapping}& $\vert \text{e}_2 \rangle, \vert\text{g}_2\rangle \leftarrow
\vert 0 \rangle, \vert1 \rangle$ & $g_{\text{g}_2\text{e}_2}$ & $\pi$ \\ \hline
\end{tabular}
\caption{Protocol for the resonant implementation of a Bogoliubov gate between qubits 1
and 2. $\Omega_R$ is the frequency of an external classical field resonant to the
transition frequency between qubit states $| \text{g}_1 \rangle$ and $| \text{e}_1
\rangle$.}
\end{figure}

As application of these resonant toolbox of two-qubit gates, we propose in the next section a protocol to implement a cluster state of $N^2$ superconducting qubits. In this cluster state, each qubit is entangled with its nearest neighbor with the help of a resonant CPHASE gate.

\section{Realization of a cluster state in a two-dimensional cavity grid}

One-way quantum computing\cite{Raussendorf01} is an alternative model to the standard
approach based on quantum gates~\cite{Nielsen04}. It requires the initial generation of a two-dimensional
cluster state, containing all necessary entanglement for the local and sequential
implementation of a quantum algorithm.  A cluster state can be obtained by the action of
a CPHASE gate between neighbouring qubits, assuming all of them initially in the
state $( \vert {\rm g} \rangle + \vert {\rm e} \rangle ) / \sqrt{2}$. Its experimental realization is one of the main difficulties when trying to implement
one-way quantum computing in different physical systems. We present here a resonant
protocol for realizing a cluster state of $N^2$ superconducting qubits. To achieve this goal, we consider a
two-dimensional cavity grid\cite{Helmer09} as the most suitable architecture in circuit QED. Indeed this architecture consists of a 2D square array of one-dimensional
resonators where the superconducting qubits are placed at the crossings of the cavities. We assume that
the qubits are initially in their ground states and the cavity-resonators contain zero photons. 

\begin{figure}[h!]
\centering
\includegraphics*[width=0.4\columnwidth]{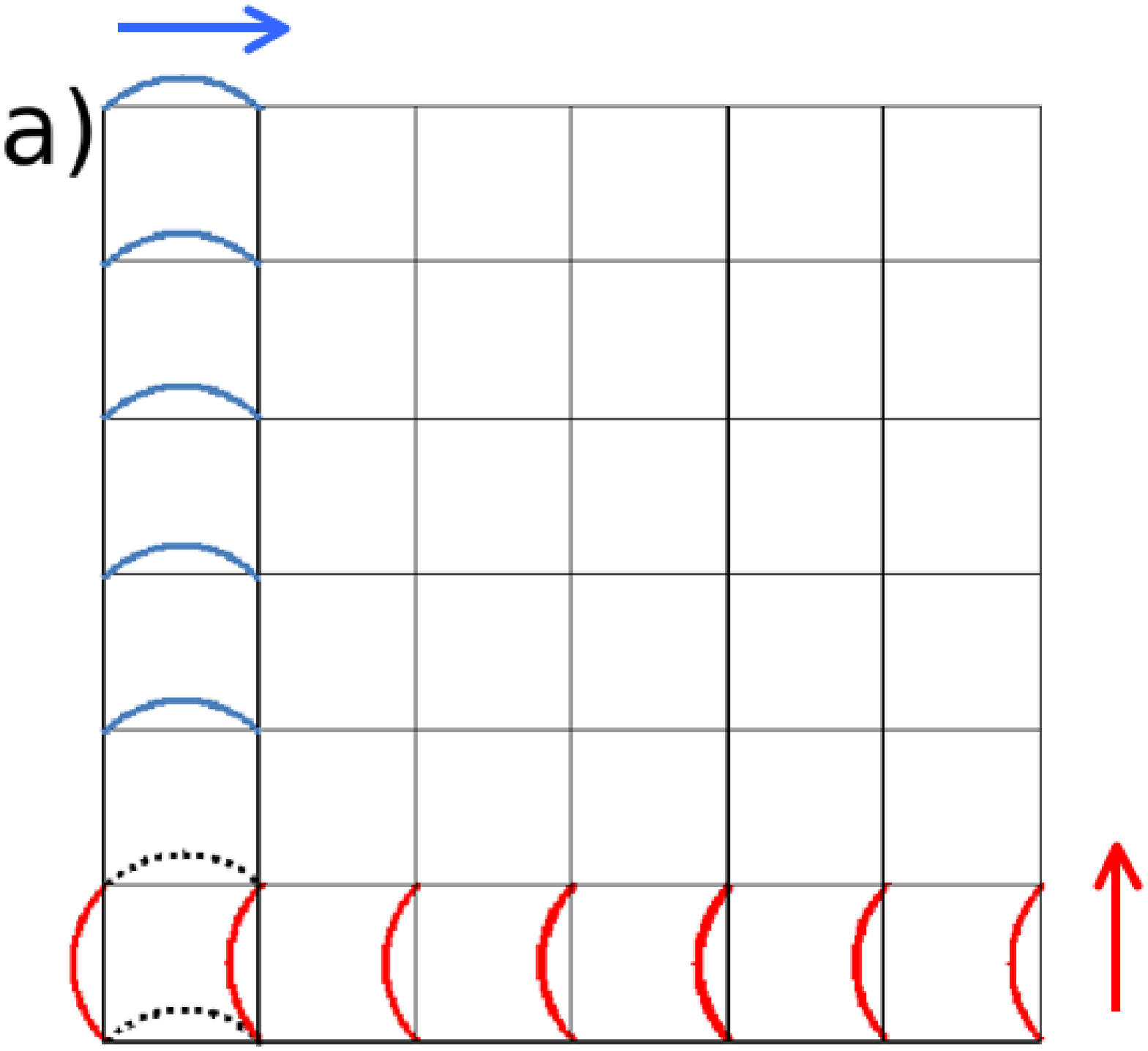}
\includegraphics*[width=0.4\columnwidth]{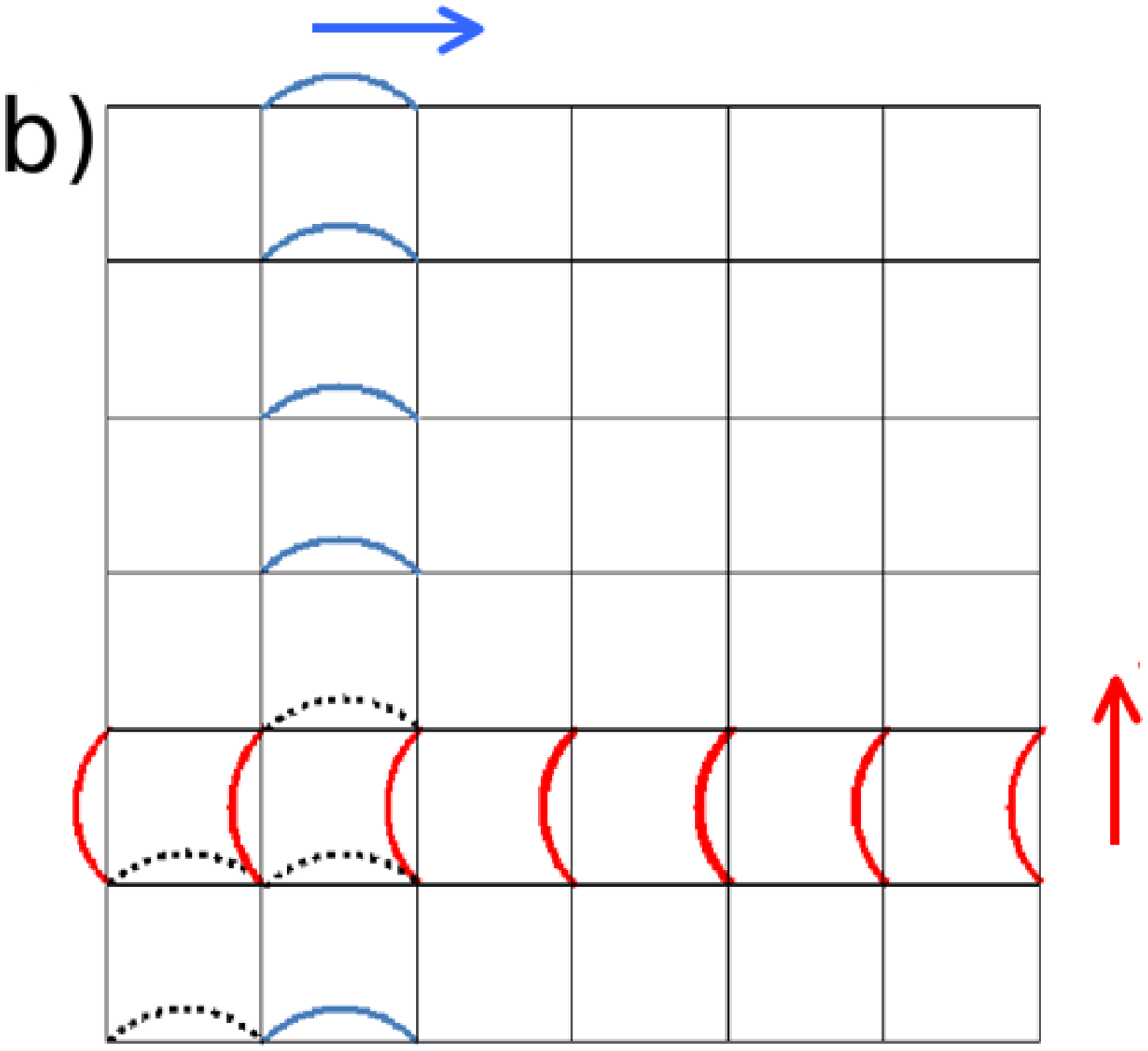}
\includegraphics*[width=0.4\columnwidth]{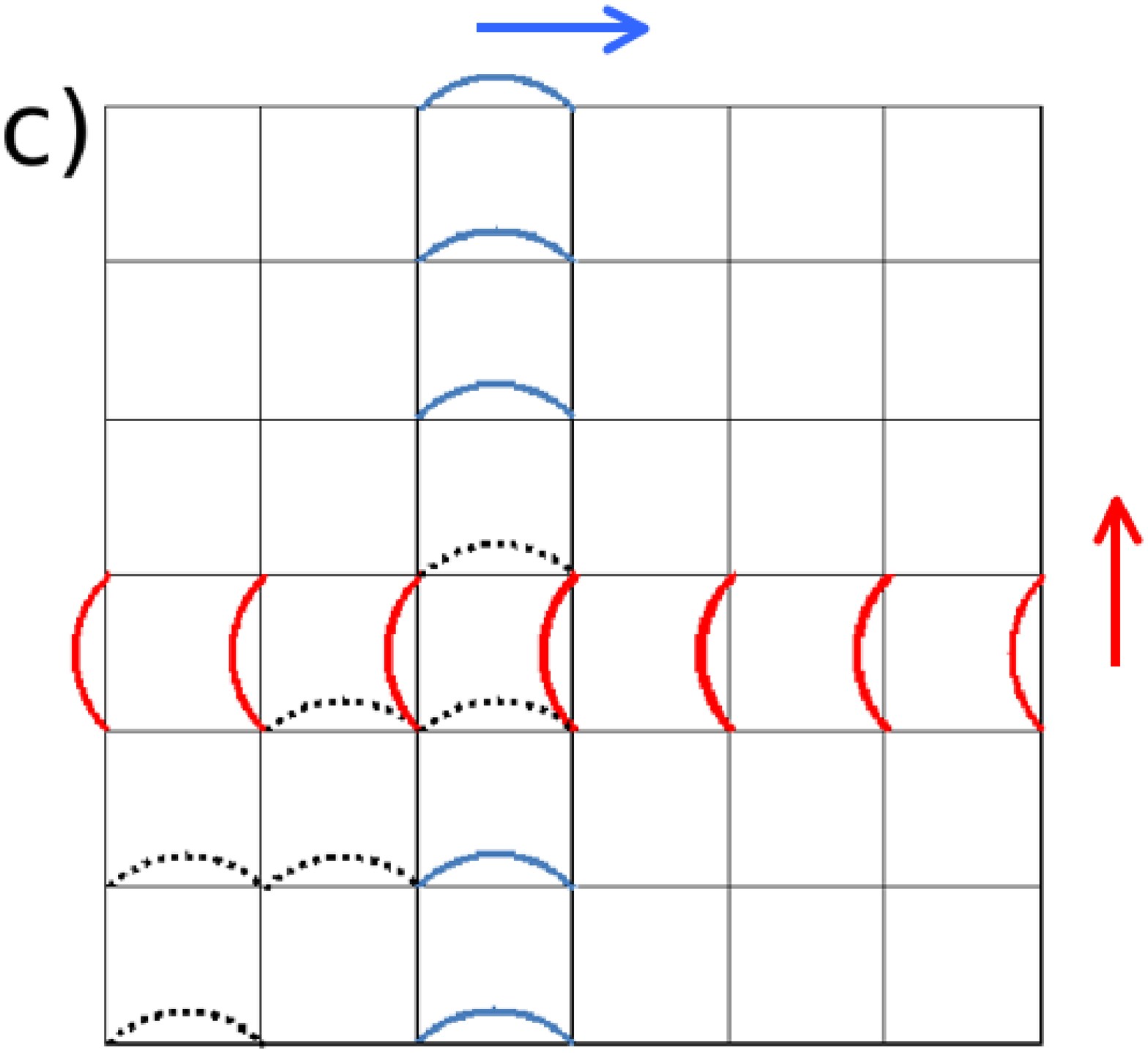}
\includegraphics*[width=0.4\columnwidth]{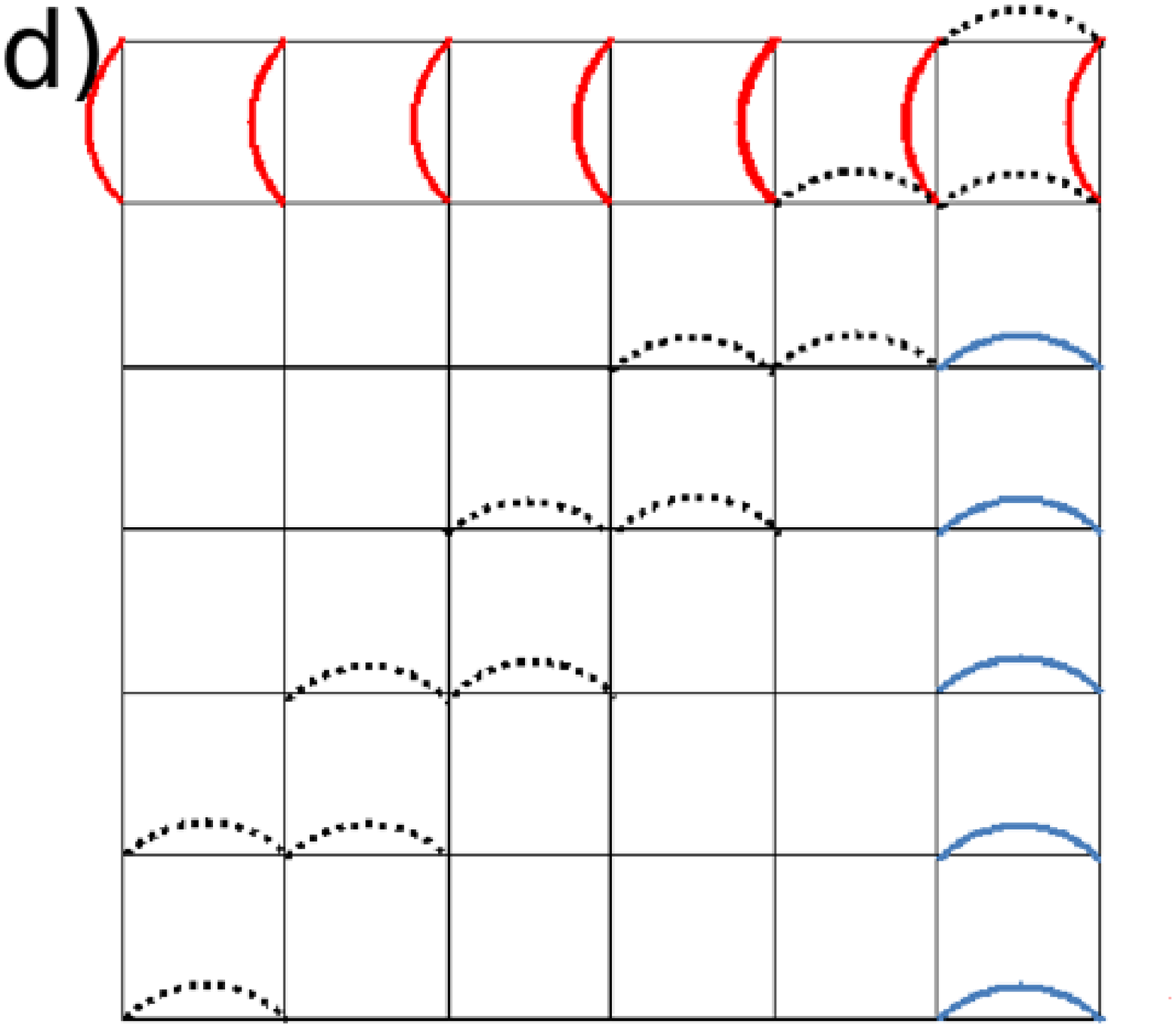}
\caption{(Color online). Generation of a cluster state in a cavity grid by implementing simultaneous CPHASE gates following steps (a)-(d). The grid is composed of
$2N$ cavities and contains $N^2$  qubits, one at each crossing point. The performed
CPHASE gates for the lines are displayed in red and for the columns in blue. The row is
shifted to the top and the column to the right. The missing gates at the crossings are
indicated as dashed black lines.}
\label{simultgates}
\end{figure}

To generate this initial state, one can send an external field resonant to the transition
frequency of all qubits to implement a y-rotation for a $\pi/2$ pulse. These local qubit rotations can be done simultaneously so that
the corresponding operation time is $\tau_{y}= \frac{\pi}{2\Omega_{R}}$, with $\Omega_{R}$ the frequency of the external field. The qubits are
then entangled by performing a CPHASE gate between each qubit and their nearest
neighbours. Although all CPHASE gates mathematically commute and could be implemented
simultaneously,\cite{Raussendorf01} the proposed architecture of the cavity grid imposes
critical constraints. In particular, one CPHASE gate can be implemented at most in one
cavity at the same time. A solution to optimize the number of steps is shown in Fig.
\ref{simultgates}. It consists in implementing simultaneously all CPHASE gates on the
first row using the vertical cavities, plus the ones we can do on the first column with
the unused horizontal cavities. We then repeat this step N times by shifting the row to
the top and the column to the right. As a result, we will have implemented all required CPHASE gates between neighbouring qubits except the ones at the crossing points between the horizontal and vertical cavities during the process. Last, we implement the
missing gates on the two diagonals in two steps. The
total time used to realize the 2D cluster state is

\begin{eqnarray}
\tau_{\rm total} = && \frac{\pi}{2\Omega_{R}} +  N \Bigg(
\frac{\pi}{g_{\text{g}_2 \text{e}_2}} + \frac{2 \pi}{g_{\text{e}_1\text{a}_1}} +
\frac{\pi}{g_{\text{g}_2 \text{e}_2}} \Bigg)  \nonumber \\
&& + 2 \Bigg( \frac{\pi}{g_{\text{g}_2 \text{e}_2}} + \frac{2
\pi}{g_{\text{e}_1\text{a}_1}} + \frac{\pi}{g_{\text{g}_2 \text{e}_2}} \Bigg) \nonumber \\
= && \frac{\pi}{2\Omega_{R}} + (N + 2) \cdot 2\pi \cdot \Bigg(
\frac{g_{\text{g}_2 \text{e}_2} + g_{\text{e}_1\text{a}_1}}{g_{\text{g}_2 \text{e}_2} \,
g_{\text{e}_1\text{a}_1}} \Bigg),
\end{eqnarray}

where the second and third term on the r.h.s. of the first line correspond to the
simultaneously done and the last missing CPHASE gates, respectively. The total time
scales with $N$, whereas the number of qubits is $N^2$, producing a resonant and
efficient technique to build 2D cluster states in the cavity grid architecture.

We consider now how errors would accumulate during this process in the case of the generation of a cluster state in a cavity grid with an array of $3 \times 3$ superconducting qubits. This example implies operating on 9 qubits, a rather advanced platform when compared with the state-of-the-art case of 3 qubits~\cite{Fink09}. For each one-qubit gate, we consider a fidelity of $99\%$, which is in agreement with recent experimental results.\cite{DiCarlo09,Chow10}. Although a fidelity of $98.5\%$ for the CPHASE gates was achieved in our numerical simulation of Section II, we consider a more conservative value of $97\%$ in our present error estimations. Following the steps presented before, we have to implement $9$ one-qubit operations and 12 CPHASE gates, reaching a global fidelity of $63 \%$. Given the number of operations, this fidelity can be considered as good for the sake of demonstrating the scheme with a small number of qubits. Clearly, full studies of decoherence sources!
  and popu

\section{Conclusion}

With this work, we extended ideas of quantum optics and cavity QED to the field of circuit QED. We considered the two-qubit CPHASE gate to introduce the tools used to implement this unitary two-qubit operation in the resonant regime and to prove its feasibility in circuit QED. Indeed, we have shown numerically that one can obtain fast resonant CPHASE gates of high fidelity, above $97\%$ by using an auxiliary qubit level and the cavity photon to establish
qubit communication. Then we have applied these tools to the implementation of other two-qubit gates, the iSWAP and Bogoliubov gates. All these resonant gates, together with high-fidelity one-qubit gates, constitute a circuit QED toolbox of resonant gates for the sake of implementing standard quantum
computing. As possible application of this resonant toolbox, we have introduced a method of generating efficiently a
cluster state in circuit QED in a two-dimensional architecture for implementations of one-way quantum
computing. We expect the introduced toolbox of resonant and efficient gates will
enhance the theoretical and experimental research in quantum information applications in
circuit QED.

\section{Acknowledgements}

The authors acknowledge valuable discussions with A. Wallraff. This work was
funded by the DFG through SFB 631 and the Nanosystems Initiative Munich (NIM) as well as the Emmy-Noether program. E.S. thanks support from UPV/EHU Grant GIU07/40, Spanish MICINN project FIS2009-12773-C02-01, EuroSQIP and SOLID European projects. G.H. acknowledges
the support of the Swiss NSF.

\end{document}